# On the physical meaning of latent track boundaries in swift heavy ion irradiated polymers

Adil Tuleushev*, Fiona Harrison, Maxim Zdorovets
*L.N. Gumilyov Eurasian National University, 010008, Astana, Kazakhstan*
* *Corresponding author. E-mail address:* adilzht@outlook.com (A.Z. Tuleushev)

**Abstract**
A large body of experimental studies of swift heavy ion latent tracks in dielectric materials has produced a wide range of estimates of track size. We investigate the physical meaning of these estimates by examining the different criteria of "track boundary" probed by various experimental techniques, including SAXS, XRD, chemical etching and conductometry. We show that different methods probe different physical aspects of ion induced modification, such as electron density redistribution, molecular ordering, chemical reactivity and charge separation, resulting in different determinations of effective track boundaries. Particular attention is paid to polymer films with electret-like properties, where post-irradiation redistribution of weakly bound electrons may play an important role in the evolution of latent track structure.

## Introduction

Cyclotron-based technologies using swift heavy ions (SHIs) for the production of nuclear track polymer membranes are well established industrial processes. Such membranes contain parallel pores with controlled diameters and surface densities and are widely used in medicine, biotechnology, and filtration technologies. Increasing membrane throughput generally involves increasing the SHI fluence applied to the polymer film. As the fluence rises, however, the probability of track overlap also increases, leading to reduced pore size uniformity and deterioration of the mechanical properties of the membrane. In practice, some degree of overlap is acceptable without significant degradation of filtration performance, although industrial specifications typically employ conservative safety margins that can unduly limit membrane throughput. Quantifying the relationship between SHI fluence and the probability of track overlap is therefore of practical importance for optimizing irradiation conditions for a given set of desired membrane parameters.

For engineering estimates of pore distribution statistics in SHI-irradiated polymer films, the Poisson distribution is commonly used [1]. Within this framework, the dimensionless parameter $FS$, defined as the product of the fluence $F$ and pore cross sectional area $S$, determines the fractions of undamaged (isolated) and overlapping pores. The fraction of membrane area free of pores is given by $P(0)=exp(-FS)$, while the fraction of isolated non overlapping pores is $P(1)=FSexp(-FS)$. The fraction of pores formed by overlapping tracks is therefore $P(overlap)=1-(1+FS)exp(-FS)$.

It has been demonstrated that chemical etching is not always required to obtain highly effective filtration membranes. In [2-4], PET films irradiated with fluences several orders of magnitude higher than those typically used in commercial membrane production exhibited exceptionally high filtration rates and pronounced ionic selectivity in aqueous electrolytes even in the absence of etching. Performance approaches that of certain biological membranes, which given the interest in the filtration properties of latent tracks, points to the need for a deeper understanding of the physical nature of latent tracks themselves.

A survey of the literature shows that there is no universally accepted definition of what constitutes a "latent track." In membrane technology, track size is usually identified with the diameter of the etched pore. Prior to etching, however, the situation is considerably less straightforward: latent tracks do not possess a sharp physical boundary, and different experimental techniques probe different types of radial modification within the irradiated material. As a result, different methods effectively define different boundaries between the latent track and the surrounding bulk material. Here, we make a comparative analysis of what different experimental methods actually measure when they assign a transverse dimension to a latent track.

## Analysis

The commercial production of PET track membranes with well calibrated pores is based on the pronounced increase, by up to three orders of magnitude, in the etching rate along the ion trajectory following additional UV treatment. The conventional interpretation is based on a model of SHI interaction with matter, according to which the latent track structure is formed through the action of fast δ-electrons and secondary electrons emitted from the track core and depositing energy in the surrounding material. This process is assumed to produce substantial chemical degradation within the core region, including extensive bond scission, density reduction, and the formation of volatile molecular fragments.

This provides an explanation of the formation of nanopore pathways facilitating penetration of the etchant and the strong anisotropy of the etching rates [5,6].

In the studies reported in [7], polycarbonate (PC), polyethylene terephthalate (PET), polyimide (PI), and polysulfone (PSU) films were irradiated with Kr ions of specific energies up to 9 MeV/u, fluxes of $(1–2)\times10^{8}$ $cm^{-2}\cdot s^{-1}$, and fluences in the range $(1\times10^{11}–6\times10^{12})$ $cm^{-2}$. Irradiation resulted in the formation of volatile products, including carbon monoxide and carbon dioxide in PET, acetylene, and, in the case of PSU, methane and ethylene. Using the Poisson-type relation $A = A_0[1-exp(-FS)]$, where $A$ is the intensity of the alkyne end group absorption band after irradiation, $A_0$ is the corresponding value for the pristine film, and $F$ is the fluence, the authors estimated an effective latent track radius in PET of approximately 4 nm at fluences around $10^{10}$ $cm^{-2}$.

Substitution of this value into the expression $P(0)=exp(-FS)$ for the undamaged fraction of the film gives $P(0)\approx0.99$ at a fluence of $1.5\times10^{10}$ $cm^{-2}$. PET films irradiated at such fluences already exhibit measurable changes in their macroscopic properties, indicating that the undamaged fraction is much smaller, suggesting that the radius determined from volatile fragment formation corresponds primarily to the innermost region of the latent track, where alkyne fragmentation occurs, while macroscopic properties such as etching behavior and molecular structural rearrangement are determined by a substantially wider region of structural and chemical modification surrounding the track core.

A particularly noteworthy observation in [7] was the appearance of sulfur-rich deposits on the surface of irradiated PSU films. The authors reported: "…we could not detect a signal from $SO_2$ or from any other sulfone derivatives, neither in the infrared nor in the mass spectra. Instead, the surface of the irradiated polysulfone foil was covered by a yellow precipitation. Using energy dispersive X-ray spectroscopy, we found an enrichment of sulfur in this surface layer." The deposition of sulfur-containing products on the film surface, together with the formation of acetylene, commonly associated with high temperature decomposition processes, was interpreted within the framework of the phenomenological two temperature thermal spike model [8–10].

According to this model, the ultrafast deposition of SHI energy into the electronic subsystem, occurring on a characteristic timescale of approximately $10^{-15}$ s, produces a transient local temperature rise through electron–phonon coupling in the vicinity of the ion trajectory. Within this model, the appearance of decomposition products of polymer chain segments that under ordinary conditions require temperatures above $10^{3}$ °C to form was considered indicative of transient thermal degradation processes within the latent track region.

The appearance of sulfur-containing deposits on the surface of irradiated PSU films is not readily explained within the framework of the known mechanisms of PSU photodegradation, for which sulfur dioxide is typically the dominant gaseous product (see, for example, [11]). This suggests that the excitation and decomposition pathways of PSU molecules under SHI irradiation differ substantially from those induced by photon excitation and may be more complex than thermal degradation.

In [12,13], latent track dimensions were estimated using small angle X-ray scattering (SAXS), with the radial electron density distribution modelled as a dense central cylinder surrounded by a Gaussian-type peripheral region. This approach yielded electron density profiles for latent tracks in PI and PC films irradiated with U ions of energies of approximately 11 MeV/u and fluences of $(2–3)\times10^{10}$ $cm^{-2}$. A pronounced deficit in residual electron density was reported within the central region of the track, extending to an estimated radius of about 3.5 nm. According to the authors, the innermost part of the latent track exhibited a reduction in electron density of more than two thirds relative to the surrounding matrix. This observation was interpreted as evidence of reduced material density within the track core. Within the framework of the δ-electron model, the density decrease was attributed to the formation of free volume associated with the removal of volatile radiolysis products.

Two remarks made by the authors are particularly noteworthy. First, this interpretation was favored because models available at the time did not predict dilatational effects associated with changes in structural ordering. Second, the measurements for both polymers were performed approximately ten years after irradiation, demonstrating the remarkable long-term stability of SHI induced latent tracks in polymer materials.

Reflecting on these remarks, we consider that there is another plausible interpretation. The reduction in electron density inferred from Xray measurements is also fully consistent with the formation of extended πconjugated systems within the central region of the latent track. It is well established that increasing π-conjugation produces a red shift of the absorption edge in SHI irradiated polymer films [14–19]. The development of extended conjugated systems is accompanied by an increase in the fraction of delocalized electrons, whose probability distribution becomes more spatially uniform over the conjugated region. Since X-ray scattering is determined by the time-averaged electron density distribution [20], increased electron delocalization produce a more diffuse and less spatially localized contribution to the scattering signal. The observed reduction in Xray scattering intensity in the track core could equally well reflect a decrease in localized electron density rather than the formation of voids or large free volume regions.

These considerations suggest an alternative interpretation of the density changes observed in the track core in [12,13]. As noted above, the preference for the "degradation and removal" mechanism was largely associated with the absence of models predicting dilatational effects related to structural ordering. The semi-amorphous structure of many polymer films, including PET, do however allow such effects to occur. In [21,22] optical studies clearly demonstrated a correlation between the degree of ordering in the amorphous phase of PET and uniaxial or biaxial stretching. Since those experiments involved samples with free surfaces, the structural rearrangements were not accompanied by significant dilatational constraints. But under localized SHI irradiation the surrounding material restricts free mechanical relaxation and increased ordering of amorphous chain segments produce local negative dilatation in the track region. The surrounding unirradiated material then acts as a constraining matrix, while densification of the amorphous phase generates free molecular volume and radial tensile stresses. Relaxation of these stresses through the gradual migration and coalescence of free volume regions toward the track center could eventually produce a reduced density central region, potentially approaching the formation of an internal free volume contour.

From this we conclude that the results reported in [12,13] are also consistent with a mechanism involving irradiation-induced structural rearrangement of the amorphous phase in PET films. We observed and investigated related effects in [23–25]. We argued that the radial redistribution of electron density produced by SHI passage, which persist for long periods due to the electret properties of PET [12,26], promotes post-irradiation ordering of mobile chain segments possessing dipole moments and hence locally increases the density of the amorphous phase. Such structural rearrangement could simultaneously generate localized free molecular volume. Within this framework, the combination of local ordering, negative dilatation, and free volume formation creates conditions for subsequent structural evolution of the latent track. At a later and slower stage of relaxation, free volume regions may gradually migrate toward the track center and coalesce, leading to the formation of a reduced density central region. The cooperative rearrangement of a substantial fraction of the molecular structure within the latent track may account for the slow kinetics of the process and its enhancement during thermal annealing. We suggested that this thermally activated collective migration of free volume may correspond to the effect observed experimentally in [24].

Additional support for this interpretation is provided by the *in situ* SAXS measurements reported in [27], which revealed an expansion of the low-density central region of the latent track during thermal annealing, accompanied by an increase in the estimated track radius. The authors interpreted these observations as evidence that the track wall in the studied samples exhibits elastic behavior. This finding is also, however, consistent with the slow post -irradiated effect of ordering of molecular structure observed in [24].

In the literature, the interaction of SHIs with matter is often discussed primarily in terms of ion energy deposition and related kinematic processes, while the role of the initial ion charge and its associated Coulomb field receive comparatively little attention. Our previous experimental studies [23–25] showed that in fact the initial charge state of SHIs has a pronounced influence on the post-irradiation structural evolution of PET films.

The emergence of a helical superstructure superimposed on the primary radial ordering pattern as neighboring latent tracks begin to overlap with increasing SHI fluence *suggests* the development of π-stacking interactions characteristic of aromatic polymers [28]. The presence of helical structures in SHI-irradiated PET films was further supported by the observed dependence of irradiation-induced optical activity on fluence and ion charge [29].

In this interpretation, the residual electric field of latent tracks originates from radial charge separation. As the SHI fluence increases, the increasing electrostatic repulsion between neighboring peripheral regions contributes to the formation of helical superstructures. Using the fluence values at which the first indications of such structures were observed, the authors of [25] estimated average intertrack distances of approximately 5 nm for $Ar^{8+}$ ions and 22 nm for $Kr^{14+}$ ions. These values may be regarded as an effective transverse size of the latent track where the track boundary is defined in terms of XRD-detectable modifications of the molecular structure. More generally, this example again illustrates that the estimated size of a latent track depends strongly on which physical or structural criterion is used to define the boundary of the track and on the experimental method employed.

In the classical description of the interaction of energetic heavy ions with matter [30], the most probable ion–electron collisions produce relatively low energy electrons with energies below the ionization potential of the medium, whereas the generation of high energy secondary δ-electrons is statistically less probable. On this basis, we proposed in [19] that SHI irradiation of PET films should result in a radial redistribution of electron density, leading to an electron-enriched core and an electron-depleted peripheral region within the latent track. This suggests an analogy with conventional redox processes, in which electron redistribution modifies the local chemical behavior of the medium. By analogy, the electron depleted peripheral region of the latent track should exhibit properties analogous to those of an oxidized material, whereas the electron enriched core should behave more like a reduced region [31]. In terms of etching behavior, this would correspond to higher resistance to etching in the track periphery and enhanced etching susceptibility near the track axis.

We further noted that, in an electret material such as PET, electron redistribution is expected to involve structural traps populated during film production rather than the covalent backbone of the polymer chains themselves. Destruction of structural traps during etching would release electrons into a relatively mobile state. Following the general concept discussed in [32], such electrons could participate in catalytic electron transfer processes that would accelerate fragmentation of ion damaged chain segments in the vicinity of the latent track core.

Numerous studies of latent track etching in polymer films have established that UV pretreatment of SHI irradiated samples substantially enhances the etching rate [33–36]. By analyzing the optical transmission spectra of PET films before and after SHI irradiation, we obtained evidence suggesting that this enhancement is related not only to the direct photochemical effects of UV irradiation on PET itself, but also to the presence of trace amounts of dicarboxystilbene (DCS), a byproduct of terephthalic acid production that forms defect units within PET chains.

DCS is a photoisomer that normally exists predominantly in the *trans* configuration. Under UV irradiation, *trans–cis* isomerization can occur, and the resulting *cis* form subsequently undergo cyclization to phenanthrene like structures upon photoexcitation [37]. The probability of such photoisomerization processes depends strongly on steric constraints, which may be reduced in the central region of the latent track [12,13], potentially favoring the formation of phenanthrene like centers near the track core. Phenanthrene derivatives are known to be chemically more reactive than the terephthalate groups of regular PET repeat units and also act as deep electron traps [38]. Destruction of such centers under alkaline etching conditions could release electrons capable of participating in reductive electron transfer reactions, facilitating fragmentation and hydrolytic cleavage of nearby ion damaged PET chain segments. Additional electron accumulation associated with UV exposure and irradiation-induced defect formation in the track core may further enhance this reductive channel, thereby contributing to the experimentally observed increase in alkaline etching rates. From this perspective, the etching rate depends not only on the chemistry of intermediate molecular fragments but also on the availability of electrons within the reaction zone. This interpretation is broadly consistent with fundamental reviews on the fragmentation of organic ions [39–42], which emphasize that aromatic and carbonyl radical anions are intrinsically unstable and often undergo fragmentation accompanied by electron emission.

In an electron-rich latent track core, hydrated electrons, whose reduction potential is approximately −2.8 V, may exhibit enhanced stability and reactivity, particularly under conditions where strong electron acceptors are absent and numerous local trapping sites are present (see, for example, [43]). Such conditions could promote longer-lived reductive electron transfer processes and successive reduction steps, thereby contributing to the experimentally observed enhancement of etching rates.

Studies in which the transverse dimensions of latent tracks are inferred from chemical etching kinetics provide additional insight into irradiation-induced structural modifications in polymers. Particularly informative in this regard is the series of works [6,44,45] devoted to latent track structure in polymer films, including PET, irradiated with Ar, Kr, Xe, Au, and U ions with specific energies of several MeV/u. In these studies, etching kinetics of SHI irradiated films were analyzed in combination with conductometric measurements, allowing reconstruction of the temporal evolution of the etched channel diameter.

The region immediately adjacent to the ion trajectory was found to exhibit an exceptionally high ratio ($>10^3$) of longitudinal to transverse etching rates. Additional UV treatment further enhanced the longitudinal etching rate in the central part of the latent track at distances up to approximately 10 nm from the track axis, which the authors associated with additional chain scission processes.

In the transverse direction, the etching behavior exhibits a complex radial profile. The etching rate is maximal in the central core region and decreases rapidly in the radial direction to the value characteristic of the bulk material. At larger distances, the etching rate decreases further, reaching a minimum at approximately 15 nm from the axis, before gradually returning to the bulk value. The authors took the position of this minimum as determining the transverse dimension of the latent track, although the surrounding halo region characterized by reduced etching rates extends to distances of nearly 100 nm from the track center.

The enhanced etching rate in the core region was attributed to extensive SHI-induced degradation of polymer molecules. To account for the reduced etching rate in the halo region, two mechanisms were proposed: (1) crosslinking of polymer chains induced by fast electrons generated during ion passage, and

(2) the influence of hydrogen produced in the central part of the track and subsequently diffusing into the periphery. Evidence for crosslinking was provided, for example, by the observation of a chloroform insoluble gel fraction in PC films irradiated with xenon ions.
Although both proposed mechanisms arise from SHI irradiation, they differ substantially in their physical nature and point to different definitions of the latent track boundary. Hydrogen diffusion continues long after the ion has passed through the material and eventually extends well beyond the region of direct ion impact. This mechanism provides a physical justification for identifying the minimum in the etching rate profile as the effective track boundary. Crosslinking, by contrast, is generated during the brief passage of the ion itself and is therefore more directly associated with the primary irradiation event. If significant crosslinking persists throughout the halo region, then this mechanism would define a considerably larger effective latent track boundary.
As noted above, XRD observations of helical superstructures emerging with increasing SHI fluence and latent track overlap can also be used to estimate an effective transverse dimension of the latent track. We note that using this mechanism to define the track boundary could lead to a time dependent value in polymer films possessing pronounced electret properties, if the post-irradiation radial charge distribution remains stable over long periods. Under these conditions, the tensile stresses arising between neighboring tracks could lead to slow cooperative rearrangement of mobile chain segments and, on longer timescales, gradual collective motion of larger macromolecular structures.
We conclude our analysis of the underlying physics of different approaches to estimating latent track dimensions by considering the polarity of the charge distribution across the track radius. The interpretation proposed in [12,13] noted that the emergence of a radial electric field follows naturally from the δ-electron model. The SAXS results were interpreted as evidence of electron depletion in the track core, implying a positively charged central region relative to the periphery, where energetic electrons thermalize and become trapped.
Our results reported in [46] suggest, however, that in PET films the charge distribution is reversed, due to the electret nature of PET and the presence of a large population of weakly bound trapped electrons [26], which migrate toward the latent track core under the influence of the SHI Coulomb field. The resulting electron influx can exceed the depletion associated with the emission of ballistic electrons, leading to the formation of an electron-enriched core and an electron-depleted periphery.
The results reported in [25] concerning the influence of SHI charge on the electron density distribution and molecular structure of PET films provide further evidence that the electric field of the SHI acts as an additional factor contributing to structural modification alongside ballistic electron processes. The Coulomb field of the SHI induces tunneling of weakly-bound electrons from trapping sites in the peripheral region of the latent track toward its core during the passage time of the SHI [47]. The authors of the latter study emphasize that, within the attosecond-level accuracy achieved in their experiment, the electron tunneling time is effectively zero, whereas the interaction time of SHIs with polymer molecules spans several hundred attoseconds. The residual charge distribution within the track will be determined by the balance between this inward electron flux and the outward flux of ballistic electrons emitted from the core region. PET films possess pronounced electret properties and contain a substantial population of trapped electrons from the film production, making the formation of a negatively charged latent track core highly plausible.

## Conclusion

The analysis presented here shows that different estimates of the radial size of a latent track are based on different criteria of physical modification. Techniques such as SAXS, XRD, chemical etching and conductometry probe different aspects of SHI-induced radial modification, including electron density redistribution, molecular ordering, chemical reactivity, and charge separation. Consequently, each experimental approach defines a different effective boundary of the latent track.
For polymer films, particularly PET, there is evidence that the electret nature of the material and the redistribution of weakly bound trapped electrons play an important role in the post-irradiation evolution of latent track structure. If so, the observed reduction in X-ray scattering intensity in the track core may be associated with the formation of extended π-conjugated systems and the resulting electron delocalization, rather than solely with large scale density loss or void formation.
These results highlight that there is no single definition of the radial size of a latent track and the underlying physical meaning of the boundary criterion depends critically on which structural, electronic, or chemical signature is probed by a given experimental method. Each measurement method provides information about some aspect of post-irradiation changes in the film, integration of which can provide a more complete understanding of latent track structure.

This research did not receive any specific grant from funding agencies in the public, commercial, or not-for-profit sectors and Dr. F.E. Harrison took part in this research as a volunteer.

**CRediT authorship contribution statement**

**A.Z.T:** conceptualization, formal analysis, writing – original draft, review. **F.E.H:** conceptualization, formal analysis, writing – original draft, review. **M.V.Z:** formal analysis, writing – original draft, review.

**Declaration of competing interest**

The authors declare that they have no known competing financial interests or personal relationships that could have appeared to influence the work reported in this paper.